\documentclass{PoS}

\def\beqn{\begin{eqnarray}} \def\eeqn{\end{eqnarray}}

\title{Higher-order and mixed QCD-QED corrections for Drell-Yan}

\ShortTitle{H.O. and mixed QCD-QED corrections for Drell-Yan}

\author{\speaker{German F. R. Sborlini}$^{\ a,b}$\\
        $^a$Instituto de F\'{\i}sica Corpuscular, Universitat de Val\`{e}ncia -- 
Consejo Superior de Investigaciones Cient\'{\i}ficas, Parc Cient\'{\i}fic, E-46980 Paterna, Valencia, Spain.\\
 		$^b$Dipartimento di Fisica, Universit\`a di Milano and INFN Sezione di Milano,
I-20133 Milan, Italy.\\
        E-mail: \email{german.sborlini@unimi.it}}

\abstract{In this article, we discuss about the Drell-Yan process focusing on the computation of radiative corrections for vector boson production. First, we describe the $q_T$-resummation formalism and its application to obtain higher-order QCD corrections. We briefly summarize the state-of-the-art in these calculations, and motivate the need of more refined theoretical predictions. After that, we center into the inclusion of mixed QCD-QED higher-order contributions, both for fixed order and resummed calculations. In particular, we present a framework based on the $q_T$-resummation formalism to simultaneously handle soft gluon and photon radiation. Finally, we discuss future extensions of this approach and its phenomenological relevance.}

\FullConference{7th Annual Conference on Large Hadron Collider Physics - LHCP2019\\
		20-25 May, 2019\\
		Puebla, Mexico}

\begin{document}

\section{History and motivation}
\label{sec:history}
The Drell-Yan (DY) process consists in the production of a vector boson (namely, $Z$, $W$ or a virtual photon) in hadronic collisions, together with its subsequent decay \cite{Drell:1970wh}. It is often considered as one of the most relevant experiments to extract highly accurate data and perform a very precise comparison with the available theoretical models. In fact, DY is used to calibrate the detectors, to test perturbative QCD, to impose constraints on PDFs and extract information about the parameters involved in the Standard Model. Moreover, from the pure theoretical side, DY can be thought as a playground to propose new beyond SM scenarios and develop novel computational techniques.

However, in order to fully exploit all the features associated to the DY process, it is crucial to improve its phenomenological description through the computation of more precise corrections. These corrections must include higher-order contributions from both QCD and the electroweak (EW) sector, since both effects starts to compete between them when going beyond the lowest orders.

In this way, the purpose of this article is to summarize recent developments pointing towards the inclusion of mixed QCD-QED corrections in predictions for DY. In Sec. \ref{sec:QCD}, we discuss some details about the techniques used to compute DY corrections, and briefly review the state-of-the-art of such computations. Then, in Sec. \ref{sec:QCDQED}, we present a discussion about the inclusion of mixed QCD-QED corrections, both for fixed order and resummed contributions. In particular, in Sec. \ref{ssec:QCDQEDresumed}, we explain the extended version of the $q_T$-resummation formalism to deal with simultaneous emissions of soft gluons and photons. The conclusions and future perspectives about this interesting topic are presented in Sec. \ref{sec:conclusions}.

\section{Computing QCD corrections}
\label{sec:QCD}
The first step to tackle the analysis of the DY process consists in establishing a proper framework to perform the computations. In this case, the presence of colliding hadrons implies that perturbation theory can not be directly applied. Thus, we rely on the factorization theorem and express the cross-section according to
\beqn
\frac{d \sigma_F}{d^2 \vec{q}_T \, dM} &=& \sum_{a,b} \int dx_1 dx_2 \, f_a^{h_1}(x_1) f_b^{h_2}(x_2) \, \frac{d \hat{\sigma}_{a+b \to F+X}}{d^2 \vec{q}_T \, dM} \, ,
\eeqn
with $d \hat{\sigma}_{a+b \to F+X}$ the partonic cross-section associated with the production of the final state $F$. This final state might include only the vector bosons (for instance, in the narrow width approximation) as well as the decay of the produced intermediate state $V^*$ into a lepton pair. In this equation, the non-perturbative information about the structure of the hadrons is contained within the parton distribution functions (PDFs), which are convoluted with the partonic cross-section (calculable in terms of a perturbative expansion in powers of $\alpha_S$). Notice that this formula allows a fully differential implementation, being $\vec{q}_T$ the transverse momentum of the produced vector boson. 

It is well known that the fixed-order expansion of the partonic cross-section fails in the low $q_T$ region due to the presence of enhanced logarithmic contributions proportional to $L = \log(M^2/q_T^2)$. The solution consists in a proper re-arrangement of the perturbative expansion which relies on the fact that $\alpha_S \, L <1 $. Thus, we can write
\beqn
\int_0^{q_T^2} dk^2 \, \frac{d\hat{\sigma}}{dk^2} &=& 1+ \alpha_S \left(c_{12} L^2 + c_{11} L + \ldots \right) + \alpha_S^2 \left(c_{24} L^4 + c_{23} L^3 + \ldots \right) + \ldots \, ,
\eeqn 
where we can appreciate that the leading logarithmic (LL) term is given by $\alpha_S^n \, L^{2n}$, the next-to-leading logarithmic (NLL) contribution is proportional to $\alpha_S^n \, L^{2n-1}$ and so on. When considering the production of a colorless final state $F$, the contribution to the $q_T$-singular part is given by \cite{Catani:2013tia}
\beqn
\nonumber \frac{d \sigma_{F+X}}{d^2 \vec{q}_T \, dM} &=& \frac{M^2}{s} \sum_{c=\{q,\bar{q},g\}} \, \left[d\sigma^{(0)}_{c \bar{c} \to F} \right] \, \int \frac{d^2 \vec{b}}{4\pi^2} \, e^{\imath \vec{b}\cdot\vec{q}_T} \, S_c(M,b) \, 
\\ &\times& \sum_{a_1,a_2} \, \int_{x_1}^1 \frac{dz_1}{z_1}\, \int_{x_2}^1 \frac{dz_2}{z_2} \, \left[H^V C_1 C_2 \right]_{a_1 a_2 \to c \bar{c}} \, f_{a_1}^{h_1}(x_1/z_1,b_0^2/b^2) f_{a_2}^{h_2}(x_2/z_2,b_0^2/b^2) \, ,
\label{eq:resummation1}
\eeqn
which involves the convolution of the partonic leading-order cross-section ($d\hat{\sigma}^{(0)}_{c \bar{c} \to F}$) with the PDFs, the Sudakov factor ($S_c$) and the hard-collinear factors (denoted symbolically as $[H^V C_1 C_2]$). The Sudakov factor depends only on the flavour of the particle which is originating the soft-collinear radiation, and it is given by
\beqn
S_c(M,b) &=& \exp \left(- \int_{b_0^2}^{b^2} \frac{dq^2}{q^2} \left[A_c(\alpha_S(q^2))\log\left(\frac{M^2}{q^2}\right) \, + \, B_c(\alpha_S(q^2))\right]\right) \, ,
\label{eq:Sudakov}
\eeqn
where the coefficients $A_c$ and $B_c$ can be computed within perturbation theory. In fact, these coefficients are related to the Altarelli-Parisi splitting functions, which allows to implement cross-checks among different computations. Notice that Eq. (\ref{eq:resummation1}) is formulated in the impact parameter space and the PDFs are evolved till the reference scale $b_0^2/b^2$. Within this formalism, the resummed contribution to the cross-section is given by
\beqn
\frac{d \hat{\sigma}_{a + b \to F}^{\rm res}}{d q_T^2}(q_T,M) &=& \frac{M}{\hat{s}} \, \int_0^\infty db \, \frac{b}{2} \, J_0(b \, q_T) {\cal W}_{ab}(b,M,\hat{s}) \, ,
\label{eq:SigmaRES}
\eeqn
where all the logarithmic terms are contained inside ${\cal W}_{ab}$. By using Mellin transformations, the $N$-moment of ${\cal W}$ is given by
\beqn
({\cal W}_{ab})_N &=& \hat{\sigma}_{a+b \to F}^{(0)} \, {\cal H}^F_N(\alpha_S(\mu_R),\mu_R,\mu_F,Q^2) \, \exp\{{\cal G}_N(\alpha_S,L,\mu_R,Q) \} \, ,
\eeqn
with $\exp\{{\cal G}_N\}$ a universal form factor and ${\cal H}^F$ the hard-collinear contribution, that contains the process-dependent terms. In the previous formula, the explicit dependence on the renormalization ($\mu_R$), factorization ($\mu_F$) and resummation ($Q$) scales is included. Both the form factor and the hard-collinear corrections can be computed within perturbation theory. In particular, the exponent ${\cal G}_N$ is obtained by using an iterative solution of the renormalization group equations (RGE) for the evolution of the QCD coupling and integrating Eq. (\ref{eq:Sudakov}); then a proper expansion is required to exhibit the logarithmic structure and identify the different contributions.

\subsection{State-of-the-art and previous calculations}
\label{ssec:stateofart}
The QCD corrections to the DY process are known since a long time. The NLO contributions were computed in the late seventies \cite{Altarelli:1978id}, whilst the NNLO corrections were obtained in the nineties and keep on being computed with different methods till nowadays \cite{Hamberg:1990np,Anastasiou:2003yy,Melnikov:2006kv,Bozzi:2008bb,Bozzi:2010xn,Boughezal:2015dva,Boughezal:2015ded}. This is due to the fact that DY constitutes a relatively simple process to play with and develop new computational techniques, as stated in the Introduction.

By applying the $q_T$-resummation formalism, it was possible to compute higher-order corrections to DY up to NNLO+NNLL accuracy. For instance, this was done through the codes \texttt{DYqT} \cite{Bozzi:2008bb,Bozzi:2010xn} and \texttt{DYRes} \cite{Catani:2015vma}, which allows to compute the inclusive $q_T$ spectrum and the fully exclusive resummed corrections, respectively. The state-of-the-art in QCD for DY is N$^3$LL+NNLO \cite{Catani:2014uta,Bizon:2018foh}, with a partial knowledge of the N$^3$LO contributions (soft-virtual approximation) and some corrections beyond the N$^3$LL accuracy.

\section{Mixed QCD-QED corrections}
\label{sec:QCDQED}
In recent times, the need for more precise theoretical predictions led to the inclusion of previously neglected effects. This is the case of EW or QED corrections because $\alpha_S^2(m_Z) \approx \alpha(m_Z)$, so NNLO QCD contributions are expected to be of the same order of magnitude as NLO EW ones. Moreover, since both theories have a very different structure, novel phenomenological effects might manifest when combining QCD and EW higher-order contributions. For instance, we have studied higher-order QED corrections to the splitting functions \cite{deFlorian:2015ujt,deFlorian:2016gvk}, and we found non-negligible charge separation effects and a crucial impact in the DGLAP equations for photon PDF evolution. Besides that, QED corrections might play an important role when studying physical processes involving photons, such as diphoton production at colliders \cite{Chiesa:2017gqx,Sborlini:2017gpl}.

In the particular case of DY, the process exhibits a high-sensitivity to tiny higher-order QCD and EW corrections because of the fact that the Born level is purely EW. Some of these EW corrections were studied in the past (see for instance, Ref. \cite{Wackeroth:1996hz}), but the combined QCD-EW effects started to be analyzed very recently. In fact, it turns out that certain channels contributing to the QCD and QED corrections to the total DY cross-section have opposite sign, which leads to an enhancement of the EW effects \cite{deFlorian:2018wcj}.

\subsection{Mixed resummation}
\label{ssec:QCDQEDresumed}
On the other hand, it is important to keep under control the corrections associated to the multiple emission of soft gluons and photons. For this purpose, we extended the $q_T$-resummation framework to deal with simultaneous QCD-QED radiation \cite{Cieri:2018sfk}. Starting from the formulation in the impact parameter space, i.e. Eq. (\ref{eq:SigmaRES}), we obtain 
\beqn
({\cal W}'_{ab})_N &=& \hat{\sigma}_{a+b \to F}^{(0)} \, {\cal H}'^F_N(\alpha_S(\mu_R),\alpha(\mu_R),\mu_R,\mu_F,Q^2) \, \exp\{{\cal G'}_N(\alpha_S,\alpha,L,\mu_R,Q^2)\} \, 
\eeqn
for the $N$-moment of the resummed structure, where ${\cal H}'^F_N$ and $\exp\{{\cal G'}_N(\alpha_S,\alpha,L,\mu_R,Q^2)\}$ are the hard-collinear and resummed factors, respectively. Both factors are expanded in power series of $\alpha_S$ and $\alpha$ according to
\beqn
{\cal H}'^F_N(\alpha_S,\alpha) &=& {\cal H}^F_N(\alpha_S) + \sum_{n=1}^{\infty} \, \left(\frac{\alpha}{\pi}\right)^n {\cal H}'^{F\, (n)}_N + \sum_{n,m=1}^{\infty} \, \left(\frac{\alpha_S}{\pi}\right)^n \left(\frac{\alpha}{\pi}\right)^m\,  {\cal H}'^{F\, (n,m)}_N \, ,
\eeqn
and
\beqn
\nonumber {\cal G}'_{N}(\alpha_S,\alpha,L) &=& {\cal G}_{N}(\alpha_S,L) \, + \,  L \, g'^{(1)}(\alpha L) + \sum_{n=2}^\infty \, \left(\frac{\alpha}{\pi}\right)^{n-2} \, g'^{(n)}(\alpha L) 
\\ &+& \sum_{n,m=1}^{\infty} \, \left(\frac{\alpha_S}{\pi}\right)^{n-2}\left(\frac{\alpha}{\pi}\right)^{m-2} g'^{(n,m)}(\alpha_S L, \alpha L)  \, ,
\eeqn
where we distinguished two kind of functions. The un-primed functions (i.e. ${\cal H}^F_N$ and ${\cal G}_N$) are the pure QCD contributions, whilst the primed ones include the QED effects. In fact, we can identify pure QED corrections and other terms that include a non-trivial mixing among QCD and QED contributions (for instance, $g'^{(m,n)}$ in the exponential factor). It is worth mentioning that, in order to obtain the logarithmic expansion of ${\cal G}$, we need to take into account the mixed evolution of the running couplings, i.e.
\beqn
\frac{d \ln{\alpha_S(\mu^2)}}{d \ln{\mu^2}} &=& - \sum_{n=0}^\infty \beta_n \, \left(\frac{\alpha_S}{\pi}\right)^{n+1} - \sum_{m=1,n=0}^\infty \beta_{n,m} \, \left(\frac{\alpha_S}{\pi}\right)^{n+1} \left(\frac{\alpha}{\pi}\right)^{m} , \
\\ \frac{d \ln{\alpha(\mu^2)}}{d \ln{\mu^2}} &=& - \sum_{n=0}^\infty \beta'_n \, \left(\frac{\alpha}{\pi}\right)^{n+1} - \sum_{m=1,n=0}^\infty \beta'_{n,m} \, \left(\frac{\alpha}{\pi}\right)^{n+1} \left(\frac{\alpha_S}{\pi}\right)^{m} , \
\eeqn
including the non-trivial mixing terms that appear beyond the leading order.

\begin{figure}[htb]
\begin{center}
\begin{tabular}{cc}
\includegraphics[width=0.46\textwidth]{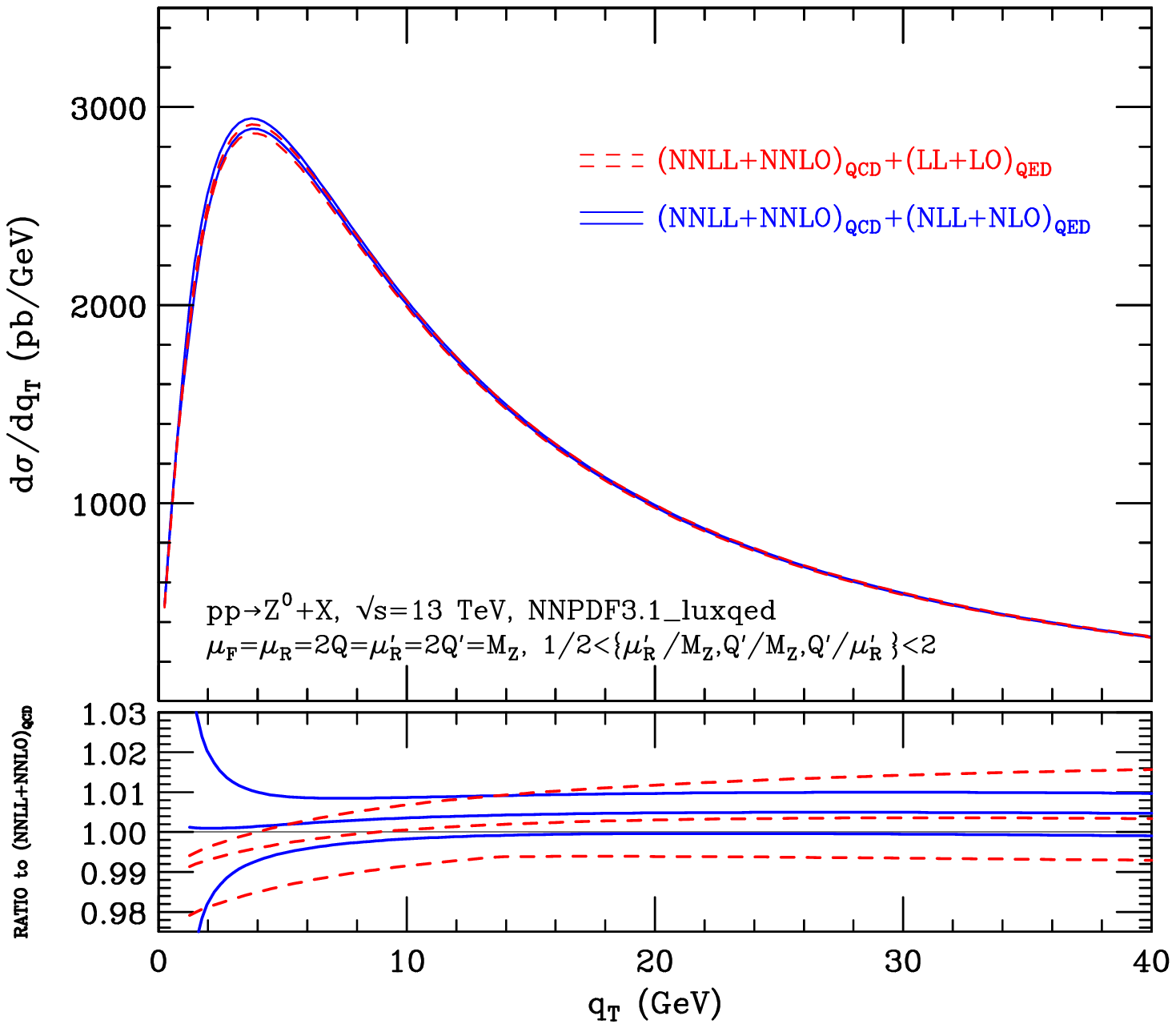} & \includegraphics[width=0.46\textwidth]{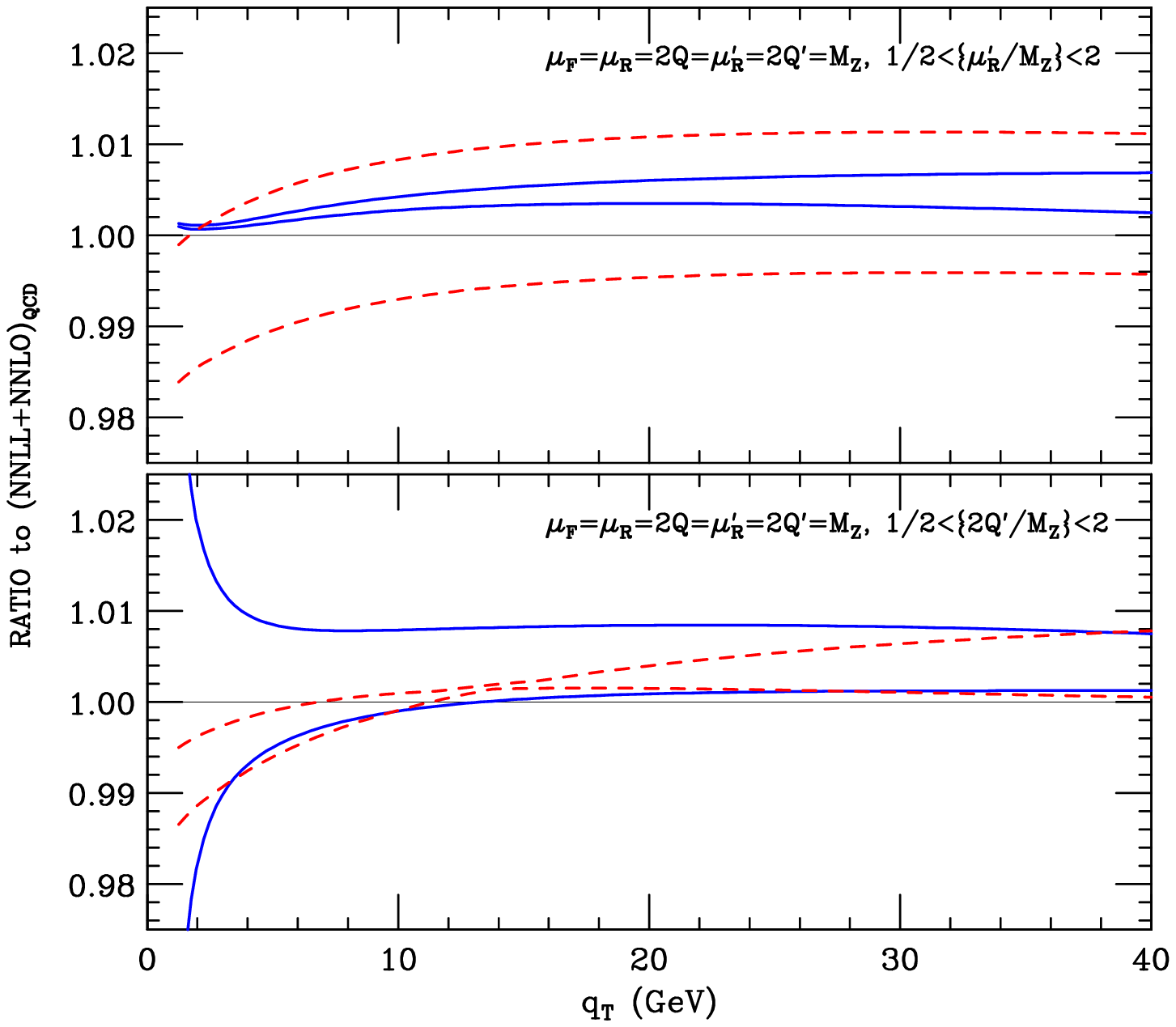}\\
\end{tabular}
\end{center}
\caption{\label{figura1}
{\em 
Inclusion of QCD-QED corrections to the $q_T$ spectrum of $Z$ boson production at LHC ($\sqrt{s}=13$~TeV). In the left panel, the NNLL+NNLO QCD results are combined with the LL (red dashed) and NLL+NLO (blue solid) QED effects. The uncertainty bands are obtained by performing a joint variation of the renormalization ($\mu_R$) and QED resummation ($Q'$) scales around their central values. We also show the ratio of the QED scale-dependent results to the standard NNLL+NNLO QCD prediction. In the right panel, we analyze the uncertainty bands when varying the resummation (upper plot) and renormalization (lower plot) QED scales. As usual, the central value corresponds to the NNLL+NNLO QCD prediction with the default scale choice. For more details, see Ref. \cite{Cieri:2018sfk}.
}}
\end{figure}

To provide a phenomenological application of the formalism, we studied the mixed QCD-QED resummation effects for $Z$ boson production at colliders. In Fig. \ref{figura1}, we show the results for LHC at 13 TeV, comparing the standard NNLL+NNLO QCD predictions (black line) and adding up to NLL+NLO QED corrections (plus the non-trivial mixing terms). Even if the corrections turn out to be small, adding these higher-order contributions reduces the scale uncertainties and makes the predictions more stable to variations in the choice of the EW scheme \cite{Cieri:2018sfk,Sborlini:2018fhr}.

\section{Outlook and conclusions}
\label{sec:conclusions}
The Drell-Yan process is a crucial experiment for particle physics, since it has many experimental and theoretical applications. Even if accurate QCD corrections have been computed since the seventies, it is necessary to include also a proper description of mixed QCD-EW contributions in order to match the quality of the measured data.

In this article, we made a brief review of the QCD higher-order corrections to DY, focusing on those obtained through the $q_T$-resummation formalism. After describing this framework, we proceeded to explain its extension to deal with simultaneous gluon/photon radiation, which allowed us to include mixed QCD-QED higher-order corrections to neutral vector boson production. In particular, we presented NLL+NLO QED corrections (also including the corresponding non-trivial QCD-QED mixing terms) for $Z$ boson production at LHC. With this, we showed that the formalism was consistent (i.e. it led to physical results) and also that the corresponding corrections to the $q_T$ spectrum are small but non-negligible (i.e. percent-level close to the resummation peak). Moreover, including these QCD-QED corrections allowed to stabilize the predictions, reducing the scale uncertainties and the dependence on the EW parameters choice. 

An extension of the formalism to deal with charged final states (i.e. $W^{\pm}$ production) is being currently investigated.

\section*{Acknowledgments}
\label{sec:Acknowledgements}
This work is supported by the Spanish Government (Agencia Estatal de Investigacion) and ERDF funds from European Commission (Grants No. FPA2017-84445-P and SEV-2014-0398), by Generalitat Valenciana (Grant No. PROMETEO/2017/053), by Consejo Superior de Investigaciones Cient\'ificas (Grant No. PIE-201750E021) and Fondazione Cariplo under the Grant No. 2015-0761.

\end{document}